**Validated Intraclass Correlation Statistics to Test Item Performance Models**

Pierre COURRIEU [a], Muriele BRAND-D'ABRESCIA [b], Ronald PEEREMAN [c],

Daniel SPIELER [d], and Arnaud REY [a]

a: LPC, CNRS, Université de Provence, Marseille, France

b: LEAD, CNRS, Université de Bourgogne, Dijon, France

c: LPNC, CNRS, Université Pierre Mendès France, Grenoble, France

d: School of Psychology, Georgia Institute of Technology, Atlanta, Georgia, USA

Running head: Expected Correlation Tests

Corresponding author:

Pierre Courrieu,

Laboratoire de Psychologie Cognitive, UMR CNRS 6146,

Université de Provence, Centre Saint Charles,

Bat. 9, Case D,

3 Place Victor Hugo,

13331 Marseille cedex 3,

France.

E-mail: pierre.courrieu@univ-provence.fr

# Validated Intraclass Correlation Statistics to Test Item Performance Models

*Abstract*. A new method, with an application program in Matlab code, is proposed for testing item performance models on empirical databases. This method uses data intraclass correlation statistics as expected correlations to which one compares simple functions of correlations between model predictions and observed item performance. The method rests on a data population model whose validity for the considered data is suitably tested, and has been verified for three behavioural measure databases. Contrarily to usual model selection criteria, this method provides an effective way of testing under-fitting and over-fitting, answering the usually neglected question "does this model suitably account for these data?"

## 1. Introduction

Theoretical models of perceptual and cognitive processing are commonly able to provide quantitative performance predictions at the item level. For instance, in the field of visual word recognition, recent models of reading are able to predict response times to individual word stimuli (the items) in various tasks, such as lexical decision, word naming or word identification (Coltheart, Rastle, Perry, Langdon, & Ziegler, 2001; Perry, Ziegler, & Zorzi, 2007, 2010; Plaut, McClelland, Seidenberg, & Patterson, 1996; Seidenberg & McClelland, 1989). Empirical databases have been collected to test these theoretical predictions at the item level, which in several cases resulted in disappointing outcomes: the tested models accounted only for a small amount of empirical item variance (Balota & Spieler, 1998; Seidenberg & Plaut, 1998; Spieler & Balota, 1997). This can result from the fact that the tested models are erroneous or incomplete, however, another possibility is that the empirical data are not accurate enough to allow good fits of plausible models, and we miss methods to clearly conclude on these two points. In fact, it has recently been shown that the amount of reproducible variance of word identification times is related to the number of participants used in the data collection by a simple law having the form of an intraclass correlation coefficient (Rey, Courrieu, Schmidt-Weigand, & Jacobs, 2009). This constitutes a suitable reference for testing item level performance models, provided that we can be sure that the considered empirical data set actually fulfils the above mentioned law. The main purpose of this paper is to provide an efficient test of this statistical model for every item level data set, to show that this statistical model actually applies to widely used behavioural measures, and to show how to use validated correlation statistics to test model predictions. This widely extends prior work of Rey et al. (2009), and provides a complete methodology to test item performance models on empirical databases.

In order to test a model, one usually collects empirical data to be compared to the corresponding model predictions, and one optimizes the free parameters (if any) of the model in order to minimize the prediction error, or to optimize some "goodness of fit" measure. At this point, one must judge (a) whether the considered model suitably accounts for the data or not, and (b) whether this model must be preferred or not to other concurrent models. Point (b) is called "model selection", and it has been widely studied in the literature (Akaike, 1974; Hannan & Quinn, 1979; Hansen & Yu, 2001; Kass & Raftery, 1995; Myung, Pitt, & Kim, 2005; Pitt & Myung, 2002; Rissanen, 1996; Schwarz, 1978). Note, however, that an answer to point (b) does not necessarily imply a similar answer to point (a), and surprisingly, this last point has been almost completely neglected in the literature, leading to difficulties in interpreting a number of results (e.g. Spieler & Balota, 1997). There are two ways for a model fit to be bad: it can be "under-fitting", or "over-fitting". Under-fitting results in a large prediction error and is generated by erroneous or incomplete models. Over-fitting is more insidious because it results in a small prediction error for the current data, but the model is not able to generalize suitably, and the results are poorly reproducible. This is a well-known consequence of using too many free parameters in a model to fit the empirical data, in such a way that the model encodes a substantial part of the data random noise instead of capturing essentially the data regularities. This is why usual model selection criteria such as the Akaike Information Criterion, abbreviated AIC (Akaike, 1974), or the Bayesian Information Criterion, abbreviated BIC (Schwarz, 1978), require optimizing a compromise between the goodness of fit (maximum log-likelihood, for these criteria) and the number of free parameters in the model. However, none of these model selection criteria allows us to detect under-fits or over-fits, they just indicate a "winner" in a given set of competing models.

Since a few years, considerable efforts have been devoted to collect and develop large scale databases that provide behavioural measures at the item level. Each item measure is usually based on an average over a number of participants. For instance, this is the case in the recent English and French Lexicon Projects (Balota, Yap, Cortese, Hutchison, Kessler, Loftis, Neely, Nelson, Simpson, & Treiman, 2007; Ferrand, New, Brysbaert, Keuleers, Bonin, Méot, Augustinova, & Pallier, 2010), which allow researchers to test various hypotheses and reading models on large sets of empirical data. Building factorial designs on such data is quite easy, however, testing item level performance models remains problematic because one does not know what can be considered as a good model fit for these data. A solution to this problem would be to provide, together with the behavioural measures, some reference "goodness of fit" measure with suitable under-fitting and over-fitting limits. This is the ultimate goal of this work, and the Matlab program named "ECVT" (for "Expected Correlation Validity Test") listed in Appendix A provides an operational solution to the problem, together with an efficient test of the validity of the adopted approach for the data set to be processed. Matlab users can directly copy the code in their Matlab editor and use it, while the listed code can also serve as an implementation model for other platforms. Comments in the code (at right of "%") provide indications on the use of the program, as well as on the actions of its various parts. The reader will also find in Appendix B an example of use of the ECVT program with helpful comments.

Hereafter, we describe the methods implemented in the ECVT program, we evaluate their efficiency and performance on artificial data, and we test their relevance on three real databases of word identification and word naming times. In Section 2, we present the adopted population model and we derive theoretical correlation functions. In Section 3, we present statistics suitable to estimate the useful correlations. In Section 4, we present a test to validate (or invalidate) the population model and derived correlations for a given data set. In Section 5, we demonstrate the efficiency and effectiveness of this test on artificial data sets. In Section 6, we validate the approach on three real behavioural databases. In Section 7, we demonstrate the use and performance of the new tool to test models. Finally, we consider recent examples of reading model fits and we conclude in Section 8.

**2. Population model**

In this section, we first define a statistical model of the behavioural measures we plan to account for. As we shall see, this is just an additive decomposition model commonly used for continuous variables (Section 2.1). From this model, we then derive a measure of the proportion of item variance that is not random, that is, the proportion of item variance that a perfect model should account for. As we shall see, this derivation results in a well-known intraclass correlation coefficient, commonly abbreviated ICC (Section 2.2). Finally, we define models' fitting measures that suitably compare to the ICC, and there are mainly two distinct kinds of models with different appropriate fitting measures (Sections 2.3 and 2.4).

**2.1. Behavioural variable model**

Let $I$ be a population of items, let $P$ be a population of participants, and let $X$ be a behavioural measure on $I \times P$ (e.g. response time). One assumes that $X$ conforms to the usual additive decomposition model:

$$X = \mu + \alpha + \beta + \varepsilon, \tag{1}$$

where $\mu$ is the mean value of $X$ on $I \times P$, and $\alpha$, $\beta$, and $\varepsilon$ are three independent random variables of mean zero, and of variance $\sigma_\alpha^2$, $\sigma_\beta^2$, and $\sigma_\varepsilon^2$, respectively. The variable $\alpha$ is the participant effect, and it takes a constant value for each given participant. The variable $\beta$ is the item effect, and it takes a constant value for each given item. The variable $\varepsilon$ is considered as a random noise, however, it can as well result from the combination of an item-participant interaction and of a true random noise. The variable $\beta$, whose values characterise the items, is the variable of interest in this study.

One can derive from $X$ another measure, denoted $X^{(n)}$, that is the arithmetic mean of $X$ over $n$ randomly selected distinct participants (thus $X^{(1)} = X$), then one obtains from (1) the following decomposition:

$$X^{(n)} = \mu + \alpha^{(n)} + \beta + \varepsilon^{(n)}, \tag{2}$$

where the random variables $\alpha^{(n)}$, $\beta$, and $\varepsilon^{(n)}$ are always independent with means zero, but their variances are now $\sigma_\alpha^2/n$, $\sigma_\beta^2$, and $\sigma_\varepsilon^2/n$, respectively.

**2.2. Item performance correlation between equal size samples of participants**

Consider now the bivariate distribution of pairs $(x,y)$, where $x$ and $y$ are independent realizations of $X^{(n)}$. Then the population correlation between $x$ and $y$, varying the items, is given by:

$$\rho(x,y) = Cov(x,y)/(Var(x)Var(y))^{1/2},$$

where, using (2), one has:

$$Cov(x,y) = Cov(\beta + \varepsilon_x^{(n)}, \beta + \varepsilon_y^{(n)}) = Var(\beta) = \sigma_\beta^2,$$

because the terms that are constant with respect to the item variable ($\mu$ and $\alpha^{(n)}$) play no role in the correlation, and the variables $\beta$, $\varepsilon_x^{(n)}$, and $\varepsilon_y^{(n)}$ are independent.

For the same reasons, one has also:

$$Var(x) = Var(\beta + \varepsilon_x^{(n)}) = Var(\beta) + Var(\varepsilon_x^{(n)}) = \sigma_\beta^2 + \sigma_\varepsilon^2/n,$$

and similarly:

$$Var(y) = Var(\beta + \varepsilon_y^{(n)}) = Var(\beta) + Var(\varepsilon_y^{(n)}) = \sigma_\beta^2 + \sigma_\varepsilon^2/n.$$

Thus, finally,

$$\rho(x,y) = \frac{\sigma_\beta^2}{\sigma_\beta^2 + \sigma_\varepsilon^2/n}. \tag{3}$$

One can recognize in (3) the expression of a well-know intraclass correlation coefficient (ICC), that is the "ICC(C, k), Cases 2 and 2A" coefficient, according to the nomenclature of McGraw and Wong (1996). To simplify the notation, it is convenient to define the ratio:

$$q = \sigma_\beta^2 / \sigma_\varepsilon^2, \quad (4)$$

so that the correlation between two independent realizations of $X^{(n)}$ is:

$$\rho = \frac{nq}{nq+1}, \quad (5)$$

which implies that:

$$q = \frac{\rho}{n(1-\rho)}, \quad (6)$$

and also that:

$$n = \frac{\rho}{q(1-\rho)}, \quad (7)$$

which are convenient formulas for finding a parameter when one knows the two other ones, usually replacing $q$ an $\rho$ by their estimates.

The ICC therefore provides, for a given dataset, a reference correlation value for model tests. As described in the following sections, a distinction has to be done however between two modelling approaches that are both designed to account for item variance. In a first approach (section 2.3), one considers theoretical item performance as generated by full simulation models able to simulate participant variability (very rare to date, but probably available in a near future), while in the second approach (section 2.4), one provides an account of theoretical item performance as generated by predictors, as in multiple regression approaches (e.g., Spieler & Balota, 1997; Yap & Balota, 2009). Note that recent simulation models are in fact used as simple predictors (e.g. Perry, Ziegler, & Zorzi, 2010).

**2.3. Item performance correlation between observed and simulated data**

Consider now a variable $V$, that could be generated, for instance, by a full simulation model, and which is affinely related to $X$ by:

$$V = aX + b \quad (8)$$

where $a \neq 0$, and $b$ are two real numbers. Then one has:

$$V^{(n)} = aX^{(n)} + b = (a\mu + b) + a\alpha^{(n)} + a\beta + a\varepsilon^{(n)}.$$

Let $x$ be a realization of $X^{(n)}$, and let $v$ be an independent realization of $V^{(n)}$. Then, there is a realization $y$ of $X^{(n)}$ such that $v = ay + b$, and:

$$\rho(x,v) = Cov(x,v)/(Var(x)Var(v))^{1/2} = aCov(x,y)/(Var(x)a^2Var(y))^{1/2} = sign(a)\rho(x,y), \quad (9)$$

and thus:

$$|\rho(x,v)| = \rho(x,y). \quad (10)$$

In other words, if a simulation model generates data that fulfil (8), then one can expect that groups of simulated participants provide mean item performance values whose absolute correlation with those of human participant groups of the same size ($n$) is given by (3)-(5).

### 2.4. Item performance correlation between a sample of participants and a predictor

Instead of building simulation models whose output fulfils (8), modellers commonly try to predict the unknown variable $\beta$ that appears in (1), without modelling the participant effect and the random variability. So, it is of interest to know what happens if a model generates a variable $B$ affinely related to $\beta$ by:

$$B = a\beta + b, \quad (11)$$

for some real numbers $a \neq 0$, and $b$. Let $x$ be a realization of $X^{(n)}$, and let $B$ be defined as in (11), then on has:

$$Cov(x,B) = Cov(\beta + \varepsilon_x^{(n)}, a\beta) = a\sigma_\beta^2,$$

$$Var(x) = Var(\beta + \varepsilon_x^{(n)}) = Var(\beta) + Var(\varepsilon_x^{(n)}) = \sigma_\beta^2 + \sigma_\varepsilon^2/n,$$

$$Var(B) = a^2\sigma_\beta^2,$$

and thus:

$$\rho(x,B) = Cov(x,B)/(Var(x)Var(B))^{1/2} = sign(a)\frac{\sigma_\beta}{\left(\sigma_\beta^2 + \sigma_\varepsilon^2/n\right)^{1/2}}, \quad (12)$$

that is:

$$\rho^2(x,B) = \rho(x,y), \quad (13)$$

where, at new, $y$ represents a realization of $X^{(n)}$ independent of $x$. In other words, if a model generates a variable that fulfils (11), then one can expect that the squared correlation of this variable with the mean item performance of a group of $n$ participants is given by (3)-(5). Note that a coefficient similar to $\rho(x,B)$ is known in the framework of the Generalizability Theory as the "Generalizability Coefficient" (Webb, Shavelson, & Haertel, 2006).

## 3. Correlation estimates

In Section 2, we defined suitable correlations at the population level. In Section 3, we present practical estimates of these correlations for finite data samples.

### 3.1 Intraclass correlation coefficient

We consider two distinct methods to estimate the ICC. The first one is based on the usual analysis of variance (ANOVA) approach. It is fast and accurate and provides reliable confidence limits for the ICC. However, it assumes that the underlying variables are approximately Gaussian. The second approach is based on a Monte-Carlo method known as "Permutation Resampling". It is distribution free and highly flexible, however, it requires much more computational effort than the ANOVA approach. We observed that the ANOVA approach is less sensitive to missing data than the Permutation Resampling approach, however, this point will not be developed in this paper.

#### 3.1.1 ANOVA approach

In practice, one randomly selects a sample of $m$ items in the item population, a sample of $n$ participants in the participant population, and data are collected in the form of an $m \times n$ matrix of behavioural measures $(x_{ij})$, $1 \le i \le m$, $1 \le j \le n$. A standard analysis of variance (ANOVA) of this matrix provides three variation sources:

1. The between rows item effect, whose mean square is denoted $MSi$, with degrees of freedom $dfi = m - 1$, and expected value $EMSi = n\sigma_\beta^2 + \sigma_\varepsilon^2$.

2. The between columns participant effect, whose mean square is denoted $MSp$, with degrees of freedom $dfp = n - 1$, and expected value $EMSp = m\sigma_\alpha^2 + \sigma_\varepsilon^2$.

3. The residual error effect, whose mean square is denoted $MSe$, with degrees of freedom $dfe = (m-1)(n-1)$, and expected value $EMSe = \sigma_\varepsilon^2$. More generally, let $N$ be the total number of available measures in the matrix, then $dfe = N - 1 - dfi - dfp$.

Then, it is easy to see that one has an estimate of the $q$ ratio (4) with:

$$\hat{q} = \frac{MSi - MSe}{n\, MSe}, \qquad (14)$$

and one can estimate the intraclass correlation coefficient $\rho(x,y)$ of (3)-(5) by:

$$\hat{\rho} = \frac{n\hat{q}}{n\hat{q}+1} = \frac{MSi - MSe}{MSi}. \tag{15}$$

Moreover, the literature provides confidence limits and test formulas for the intraclass correlation coefficient (McGraw & Wong, 1996; Shrout & Fleiss, 1979). The confidence interval of probability $1-\alpha$ of (15) is given by:

$$\left[1 - \frac{F_{1-\alpha/2}(dfi, dfe)}{F_{obs}}, \quad 1 - \frac{1}{F_{obs} \times F_{1-\alpha/2}(dfe, dfi)}\right], \tag{16}$$

where $F_{obs} = MSi/MSe$, and $F_p(a,b)$ is the quantile of probability $p$ of Fisher $F$ distribution with $a$ (numerator) and $b$ (denominator) degrees of freedom. Take care to the reversed order of degrees of freedom for the upper confidence limit in (16). Note also that, in this context, $\alpha$ denotes the usual type I error risk (not the participant effect).

Special approaches of the intraclass correlation have been developed for the particular case of binary observations (Ahmed & Shoukri, 2010), which can be useful for the analysis of accuracy variables, for instance. However, in this paper, we more particularly focus on continuous behavioural variables such as reaction times.

### 3.1.2 Permutation Resampling approach

The analysis stated in Section 2.2 clearly shows the relation between the intraclass correlation and the correlation of average vectors. This suggests the possible use of a Monte-Carlo type method named "Permutation Resampling" (Opdyke, 2003) to compute the intraclass correlation coefficient. Despite the computational effort this method requires, Rey et al. (2009) preferred it because it is distribution free. Another advantage is the flexibility of this method in what concerns the number of participants taken into account, which will allow us to build a useful test in Section 4 below.

The Permutation Resampling procedure is as follows. Given a data table of $m$ items × $n$ participants, first choose a group size $n_g \leq n/2$. Then randomly sample two independent groups of $n_g$ participants each, calculate item means for each group, and compute the correlation coefficient $r$ between the two resulting vectors of size $m$. Repeat this $T$ times, then the average of obtained $r$ values is an estimate of the intraclass correlation coefficient (ICC) for a data set of $n_g$ participants. The larger is $T$, and the more accurate is the estimate.

In order to obtain the ICC for the whole data set with $n$ participants, one uses the average correlation and $n_g$ to obtain an estimate of $q$ by (6), then one extrapolate the desired ICC using (5) with $q$ and $n$ as arguments.

**3.2 Model correlation**

There are two cases that must be distinguished, the case of full simulation models (section 2.3), and the case of predictors (section 2.4). In both cases, human data are summarized in the form of a $m$ components vector of mean item performances:

$$x_i = \frac{1}{n}\sum_{j=1}^{n} x_{ij}, \quad i = 1...m . \tag{17}$$

In the case of a full simulation model, the model prediction vector is of the form:

$$v_i = \frac{1}{n}\sum_{j=1}^{n} v_{ij}, \quad i = 1...m , \tag{18}$$

and if the model data fulfil (8), then one has the null hypothesis (10), where the estimate $\hat{\rho}$ of $\rho(x,y)$ is given by (15), and the estimate of $\rho(x,v)$ is the Pearson $r$ correlation statistic between the vectors (17) and (18).

In the case of a simple predictor, this one is of the form:

$$B = (b_i), \quad i = 1...m , \tag{19}$$

and if it fulfils (11), then one has the null hypothesis (13), where the estimate $\hat{\rho}$ of $\rho(x,y)$ is given by (15), and the estimate of $\rho(x,B)$ is the Pearson $r$ correlation statistic between the vectors (17) and (19).

In both cases, the model fit statistic is a powered absolute correlation of the form $|r|^c$, $c \in \{1, 2\}$, with $c = 1$ for simulation models, and $c = 2$ for predictors. Under the null hypothesis (10) or (13), $|r|^c$ must belong to the ICC confidence interval (16) with probability $1 - \alpha$ of this interval. If it does not, then one can reject the null hypothesis (with risk $\alpha$), and conclude that the considered model does not suitably fit the data. Given that the ICC is the reference correlation value that model fit statistics must match as closely as possible, we refer to the ICC as the "Expected Correlation", in this context.

**4. Expected correlation validation test**

The validity of the approach developed in Sections 2 and 3 critically depends on the assumption that the considered behavioural measure fulfils the additive decomposition model (1), or an equivalent variant, which leads to the law (3) for the expected correlation. However, this is not necessarily the case for every experimental variable, and thus, a prior condition to the use of an expected correlation like the ICC (15), as a reference value to test models, is that one can verify that the considered data actually fulfil the law (3). In order to do this for their word identification time database, Rey at al. (2009) used a series of Permutation Resampling procedures, like the one described in Section 3.1.2, with distinct participant group sizes ($n_g$'s). Then they computed an estimate of the $q$ ratio that minimized the sum of squared differences between the observed ICC estimates and those predicted by (5) for the various selected $n_g$ values. The predicted and observed ICCs, as functions of the group size, were plotted in order to allow visual comparison, and the similarity of the two graphs appeared impressive, leading to the conclusion that the data suitably fulfilled the expected correlation model (3). The conclusion was correct in this case, however, visual appreciation is not always easy and reliable, as will be shown below. Another available information is the prediction error measure, however, we do not know the critical error magnitude (if any) to reject the model (3) for the considered data. So, we need a clear and easy to use test of validity of the expected correlation model (3) for every item level data set. In fact, such a test can easily be built using a procedure similar to the one described above, but where one replaces the prediction error measure by a suitable statistic whose theoretical distribution is known.

Consider the empirical distribution of $T$ correlation values generated by Permutation Resampling for a given group size $n_g$ (see Section 3.1.2). This distribution has an average $\bar{r}_g$, which is possibly an estimate of the ICC for $n_g$ participants, and its variance is denoted $s_g^2$. Let $\rho_g$ be the true, unknown, expected correlation for group size $n_g$. Given that the sampled correlation values are independent realizations of the same bounded random variable (in [-1, 1]), all moments of this variable exist, and the Central Limit Theorem does apply. Thus, as $T$ increases, the average correlation $\bar{r}_g$ rapidly converges to a normally distributed random variable of mean $\rho_g$, and of variance $s_g^2/T$. This implies that the random variable $T^{1/2}(\bar{r}_g - \rho_g)/s_g$ is normally distributed with mean 0 and variance 1. Now, consider a series of $K$ independent Permutation Resampling estimations, for $K$ distinct group sizes, then, by definition of the $\chi^2$ random variable with $K$ degrees of freedom, one has:

$$\sum_{g=1}^{K} \left( T^{1/2} (\bar{r}_g - \rho_g) / s_g \right)^2 \to \chi^2(K). \tag{20}$$

If one hypothesises that (3) is valid for the considered data set, then there is a constant $q$ such that, by (5), one has the null hypothesis:

$$\rho_g = \frac{n_g q}{n_g q + 1}, \quad g = 1..K. \tag{21}$$

The optimal determination of $q$ is the one that minimizes (20), while the $\rho_g$'s in (20) are determined by (21). In practice, this is easy to obtain using a local search procedure such as Newton-Raphson iterations for zeroing the derivative of (20) with respect to $q$. This is implemented in the Matlab sub-function named "minChi2" listed in Appendix A. In the ECVT program, one uses $T = 500$, which was found to provide accurate results with an acceptable computational effort.

The choice of the series of $K$ group sizes is somewhat arbitrary, and it is partially constrained by the total number of available participants ($n$). In the Matlab program listed in Appendix A, the series are built in order to obtain $K$ equally spaced group sizes, while $K$ is as close as possible to 12, the greatest group size is equal to the greatest integer lower or equal to $n/2$, and the lowest group size is minimally greater than or equal to the group size spacing. Note that using a very small group size can cause resampling difficulties in cases where there is a certain amount of missing data in the data set.

Finally, if the $\chi^2(K)$ value, obtained by (20) in the conditions described above, is significant, then the null hypothesis (21) can be rejected (with the chosen risk), which means that (1) and (3) probably do not provide a valid model for the considered data.

**5. Testing the test**

In order to examine the performance of the test described in Section 4, we are going to test artificial data sets that fulfil or do not fulfil the variable model (1), by construction. In order to have an idea of the discrimination power of the test, it is desirable that the data can deviate from (1) at various degrees, including a degree zero, which is simply the conformity to (1). This can be obtained by generalizing (1) in the following way:

$$X = \mu + \alpha + \gamma\lambda + \varepsilon, \tag{22}$$

where $\mu$, $\alpha$, and $\varepsilon$ are defined as in (1), $\lambda$ is the normalized item effect with mean 0 and variance 1, and $\gamma$ is the "participant sensitivity" to the item effect. The participant sensitivity has a fixed value for each participant (as $\alpha$), and has global mean $\bar{\gamma}$ and variance $\sigma_\gamma^2$. In the special case where $\sigma_\gamma^2 = 0$, one obtains (1), with $\beta = \bar{\gamma}\lambda$. For the generalized model, it is convenient to define the two following ratios:

$$q = \bar{\gamma}^2 / \sigma_\varepsilon^2, \tag{23}$$

$$u = \sigma_\gamma^2 / \sigma_\varepsilon^2. \tag{24}$$

Thus, (22) reduces to (1) iff $u = 0$, and, as one can verify, (3) is valid only in this case.

Artificial data tables, of 360 items by 120 participants, have been generated using (22) with $q = 1/16$, and four values of $u \in \{0, 1/36, 1/16, 1/4\}$. Figure 1 shows four plots generated by the Matlab function "ECVT" listed in Appendix. Each plot compares the series of observed $\bar{r}_g$ values with the corresponding $\rho_g$ values predicted by (21), for a given value of $u$, and the result of the validity test (20) appears in the title of the plot. As one can see, the two graphs are confounded, and the test is clearly non significant for $u = 0$. However, for all non-zero values of $u$, the test is highly significant, and thus, the non-conformity of the data to model (1) is detected. Moreover, one can observe that for $u = 1/36$, the test detected the existing difference, while this one is not visible at the ordinary figure scale. In fact, a very small difference becomes visible when the figure is enlarged. This not only suggests that the test (20) is powerful, but also that visual inspection of graphs is not reliable enough in this problem. The four experiments of Figure 1 were repeated 200 times each, and one recorded the frequency of rejection of the null hypothesis for two conventional type I error risks ($\alpha = 0.01$, and $\alpha = 0.05$), and for each value of $u$. As one can see in Table 1, the frequency of rejection with $u = 0$ is close to the chosen $\alpha$ risk, as expected. The frequency of rejection is very high with $u = 1/36$, and it is the maximum possible for the two greatest values of $u$. So, the validity test (20) is visibly efficient and we can use it on real data.

Insert Table 1 about here

Insert Figure 1 about here

## 6. Testing real response time databases

Tests on artificial data allowed us to be sure that our tools suitably work. Now, a crucial question is to know whether or not the proposed statistical model actually applies to real behavioural data. We examine this question hereafter on three real reaction time databases, involving two word reading tasks (word identification and word naming), and two languages (English and French).

### 6.1 Word identification times from Rey et al. (2009)

This database and methodology details are described in Rey et al. (2009). It is a set of 120 items by 140 participants word identification times, with about 4% missing data. The stimuli were 120 monosyllabic, five-letter English printed words, randomly selected in the Celex lexical database (Baayen, Piepenbrock, & van Rijn, 1993). The used task was a standard perceptual identification in a luminance-increasing paradigm (Rey, Jacobs, Schmidt-Weigand, & Ziegler, 1998; Rey & Schiller, 2005). Participants were undergraduate students at Arizona State University, native English speakers, with normal or corrected-to-normal vision.

The data table was given as argument to the Matlab function ECVT listed in Appendix A. The output provided an overall ICC equal to 0.9016, with a 99% confidence interval of [0.8655, 0.9315]. The correlation fit plot is shown in Figure 2, and the test (20) is clearly non-significant ($\chi^2(14)=8.62$, n.s.). Thus, the correlation model (3) suitably accounts for these data, and the ICC above is a reliable expected correlation to test models.

Insert Figure 2 about here

### 6.2 English word naming time

*Participants*. Ninety-four undergraduate students from Stanford University participated in the experiment. All participants were native English speakers with normal or corrected-to-normal vision.

*Stimuli and apparatus*. 770 English disyllabic words randomly selected from the Celex Database were used. The words were four-to-eight letter long, without plural forms.

*Procedure*. Each trial started with a fixation point that was presented for 500 ms on a PC computer screen. It was immediately followed by a word that appeared in the middle of the computer screen in font Courier 24. The word remained on the screen until the participant's response. Participants were instructed to read aloud the target word as quickly and accurately as possible. The interval between trials was 1,500 ms. Response times were measured from target onset to the participant's response. The experimenter sat behind the participant and recorded errors and voice key failures. The experiment started with a training session composed of ten trials. The experiment then started with test words presented in a randomized order for each participant with a break every 150 trials.

The resulting database is a set of 770 items by 94 participants word naming times, with 3.61% missing data. The data table was given as argument to the Matlab function ECVT listed in Appendix A. The output provided an overall ICC equal to 0.9261, with a 99% confidence interval of [0.9160, 0.9355]. The correlation fit plot is shown in Figure 3, and the test (20) is clearly non-significant ($\chi^2(11)=10.35$, n.s.). Thus, the correlation model (3) suitably accounts for these data, and the ICC above is a reliable expected correlation to test models. Details of the above analysis are listed in Appendix B as an example of use of the ECVT program, with helpful comments.

<u>Insert Figure 3 about here</u>

**6.3 French word naming time**

*Participants*. One hundred undergraduate students from the University of Bourgogne participated in this experiment. All were native French speakers with normal or corrected-to-normal vision.

*Stimuli*. A list of 615 French disyllabic words randomly selected from the Brulex Database (Content, Mousty & Radeau, 1990) was used. The selection was restricted to four-to-eight letter words and excluded verbs and plural forms.

*Procedure*. The same procedure as the one in the English word naming experiment (Section 6.2) was used.

The resulting database is a set of 615 items by 100 participants word naming times, with 3.94% missing data. The data table was given as argument to the Matlab function ECVT listed in Appendix A. The output provided an overall ICC equal to 0.9578, with a 99% confidence interval of [0.9513, 0.9638]. The correlation fit plot is shown in Figure 4, and the test (20) is clearly non-significant ($\chi^2(12)=6.60$, n.s.). Thus, the correlation model (3) suitably accounts for these data, and the ICC above is a reliable expected correlation to test models.

Insert Figure 4 about here

As a conclusion to Section 6, we note that all the examined real data sets seem to fulfil the variable model (1), and the resulting correlation model (3). This is not a trivial result, since the generalized variable model (22), with a non-constant "participant sensitivity" to the item effect, can a priori seem more plausible. Fortunately, the obtained results show that very commonly used behavioural measures such as word identification and word naming times can be analysed in terms of the restrictive model (1), and thus, the methodology derived from (1) to test simulation or regression models can be applied to these variables.

## 7. Testing regression models on simulated data

### 7.1. Simulated Data

In order to build a test problem, one first chooses the number $m$ of items, the number $n$ of participants, and the exact number of parameters $k_0$ that the data-generating model will use for generating its regular part. One also chooses $k_{max}$, the maximum number of free parameters that tested models can use to fit the data. One must have the inequalities: $1 < k_0 < k_{max} \leq (k_0 + n) < m$. In addition, one chooses the noise standard deviation ($\sigma_\varepsilon$), which allows approximate control of the data $q$ ratio.

One uses (1) to generate an $m$ items by $n$ participants data sample matrix ($x_{ij}$) in the following way:

$$x_{ij} = \mu' + \alpha_j + \beta_i + \varepsilon_{ij}, \quad 1 \leq i \leq m, \quad 1 \leq j \leq n.$$

The sample mean ($\mu'$) is a random constant. The sample participant effect ($\alpha_j$) is a random Gaussian vector of length $n$, with zero mean and unit standard deviation. The sample item effect ($\beta_i$) is a random Gaussian vector of length $m$, with zero mean and unit standard deviation. The sample noise is a random Gaussian $m \times n$ matrix $E = (\varepsilon_{ij})$, with mean zero and standard deviation $\sigma_\varepsilon$. Each column of the noise matrix is orthogonal to the item effect vector, however, the noise matrix itself is not orthogonal.

In order to build a base for regression models of "predictor" type (with $c = 2$, see Section 3.2), one first builds a $m \times k_0$ orthogonal matrix $H = (h_{ij})$ whose first column vector has $m$ equal components ($1/m^{1/2}$), and the remaining $k_0 - 1$ columns are orthogonal Gaussian random vectors of length $m$, with zero means and unit norms, whose sum is proportional to the sample item effect, more precisely:

$$\frac{(m-1)^{1/2}}{(k_0-1)^{1/2}} \sum_{j=2}^{k_0} h_{ij} = \beta_i, \quad 1 \leq i \leq m.$$

A predictor with $k$ degrees of freedom (free parameters) uses a base $G_k$ made of the $k$ first columns of $H$ if $k \leq k_0$, to which one adds the $k - k_0$ first columns of $E$ if $k > k_0$, so $G_k$ is a $m \times k$ matrix, and the predictor parameters ($w$ vector) are optimized as a least-squares solution of the equation $G_k w = x$, where $x = (x_i)$ is the mean item performance vector given by (17). Thus one has $w = G_k^\dagger x$, and the predictor is $B_k = G_k w$. Observe that the exact predictor can be obtained only with $k = k_0$. If $k < k_0$, then the predictor under-fits the data. If $k > k_0$, then the predictor over-fits the data.

**7.2 Under-fit and over-fit detection using the expected correlation**

The method described in Section 7.2 was used to built artificial problems with the parameter values $n = 40$, $k_0 = 20$, $k_{max} = 60$, $m \in \{61, 610\}$, and two levels of $q$ approximately equal to $1/4$ and $1/16$, respectively. In each problem, 59 models whose complexity varied from 2 to 60 free parameters were fitted to the data by the least-squares method, and one computed the squared correlation of each model prediction vector with the data average vector. Figure 5 shows the variation of the squared correlation as a function of the model complexity for the two levels of $m$ and of $q$. Also is shown in each plot the expected correlation (ICC) and its 99% confidence interval. Note that the squared correlation always intersects the ICC confidence interval in the neighbourhood of the exact complexity (20 parameters). Squared correlations under the lower confidence limit are detected as under-fits, and squared correlations above the upper confidence limit are detected as over-fits. The four experiments of Figure 5 were repeated 200 times each, in order to observe the frequency of under-fit and over-fit detections as a function of model complexity. The results are shown in Figure 6, where one can see that the minimum global frequency of misfit detections is always on a close neighbourhood of the exact complexity level (20 parameters). The under-fit detection frequency rapidly increases as the model complexity decreases from the optimum, while the over-fit detection frequency more gradually increases as the model complexity exceed the optimum. The accuracy of misfit detections increases as $m$ increases, and it is moderately sensitive to the $q$ ratio. Table 2 shows the detail of the frequency of abusive detections of under-fits and over-fits at the exact model complexity (20 parameters). For $m = 610$, this frequency is exactly the expected one, given the $\alpha$ risk (0.01). For $m = 61$, the misfit detection frequency is a bit greater than expected, however, the discrepancy is small enough to allow practical use, provided that one uses $\alpha$ risks not greater than 0.01.

<u>Insert Table 2 about here</u>
<u>Insert Figure 5 about here</u>
<u>Insert Figure 6 about here</u>

## 8. Discussion and conclusion

We have shown that, provided that the considered behavioural variable fulfils the usual decomposition model (1), one can build a suitable reference correlation (or "expected correlation") having the form of an intraclass correlation coefficient. The lower and upper confidence limits of this ICC can be considered as under-fitting and over-fitting limits, respectively, for model goodness of fit statistics, which are the absolute correlation (for full simulation models), or the squared correlation (for predictors) of model item performance predictions with empirical data averaged over participants. We demonstrated the effectiveness of this approach on artificial data that, by construction, fulfilled the variable decomposition model (1). In order to verify that any given data set fulfils model (1), and thus that the above methodology is suitable for these data, we proposed a test which is able to detect even weak deviances to this model. The performance of this test has been demonstrated on artificial data whose deviance to model (1) was gradually varied. Moreover, we tested real behavioural data sets in order to have an idea of the realism of model (1), and of the suitability of the derived methodology in practice. Three databases were tested: one set of English word identification times (from Rey et al., 2009), one new set of English word naming times, and one new set of French word naming times. It turned out that these three databases were compatible with model (1), demonstrating that the proposed methodology has a wide potential application field. Finally, the Matlab program "ECVT" listed in Appendix A allows Matlab users to directly apply this methodology, while it can also serve as an implementation model for developers on other platforms. In addition, Appendix B provides a commented example of use of the ECVT program with the data of Section 6.2.

As an ultimate illustration, let us consider two word reading model fits recently published in the literature. The models are 1) a multiple regression model with many predictors, which was used by Yap and Balota (2009) to predict word naming latencies, and 2) a simulation model (CDP++) published by Perry, Ziegler, and Zorzi (2010), which was used as a simple predictor for data similar to those of Yap and Balota, that is, a subset of the word naming latencies from the ELP database (Balota et al., 2007), corresponding to more than 6000 monomorphemic multisyllabic English words. On these data, Yap and Balota obtained a global fit of $R^2 = 0.612$, while Perry et al. obtained a global fit of $R^2 = 0.494$ after combining CDP++ predictions with usual phonological onset factors.

What can we say about the performance of these models? First, these are currently the best known fits for these two types of models on such data. But are these fits good? On one side, Yap and Balota's result seems better, however, a multiple regression model with many predictors has always an important risk of over-fitting. If one assumes that the ICC of the data set is about 0.5, for instance, then Perry et al. model could be the best. Unfortunately, one does not know the ICC of the used data set. However, it is possible to compute an estimation of its order of magnitude.

Firstly, one can reasonably assume that the items used in Section 6.2 (bisyllabic English words) are a random sample of items belonging to the same item population as those used to test the above models. Secondly, there is no reason to think that the participant populations are basically different (American college students). Thirdly, note that the expected ICC strongly depends on the number of participants (Equations 3-5), but not on the number of items. In fact, the number of items plays an important role only for the variance of ICC estimators, not for their expected magnitude. The remaining critical element is the $q$ ratio (Equation 4), which can vary depending on the conditions in which the data were collected (noise). So, we clearly take a risk assuming that the $q$ ratio of the ELP database and the one of the database of Section 6.2 are comparable. With this caution in mind, we can attempt to approximate the ICC of Yap and Balota data using our $q$ ratio for the English word naming times ($q = 0.1333$, see Appendix B), and applying Equation 5 with $n = 25$ participants (which is the number of observations per item for the naming data in ELP). Doing this, one obtains an ICC of about 0.769. Clearly, none of the above models reaches such a fitting level, indicating that the race for new reading models remains open. However, note also that a firm conclusion on this point cannot be drawn as long as one does not know the ICCs of data sets on which models were tested.

As a conclusion, it appears desirable to encourage the use of statistics like the ones presented in this paper, or possible equivalent, in order to allow researchers involved in modelling to have a clear idea of "how far they are from the truth" (the truth of the data of course!) when they test their models. Comparing the performance of various models is probably useful but clearly not sufficient. Having a quite precise idea of the distance from the target result is a precious information which can considerably help modellers improving the models. If the fit is quite close to the data ICC, probably minor changes in the model or a simple parameter tuning are sufficient. If the fit is far from the ICC, more important changes are probably necessary. If the model over-fits the data, then one must reduce the number of degrees of freedom of the model. But without a reliable reference fit, such as the data ICC and its confidence limits, the target result is not defined.

# Appendix A

## Matlab code (version 7.5) of the ECVT program

```
function [qAV,icc,conf,r,Chi2,Chi2df,Chi2p,mitem] = ECVT(x,tf,miss,pconf)
% ----------------Expected Correlation Validity Test----------------------
%                                 input:
% x:   (m items) X (n participants) data table
% tf: title of correlation fit plot (default: tf = '' for no figure)
% miss:  numerical code for missing data in table x (default = inf)
% pconf: probabilities of ICC confidence intervals (def. [.95 .99 .999])
%                                 output:
% qAV,icc: q ratio and intraclass correlation (ICC) of table x by ANOVA
% conf:   ICC confidence intervals ([probability lower upper])
% r: estimate of the ICC by Permutation Resampling and extrapolation
% Chi2,Chi2df,Chi2p: correlation validity test (Chi^2, d.f., p), where the
%    expected correlation (ICC) is not reliable if the test is significant.
% mitem: (m X 1) column vector of mean performance for each item.
% -----------------------------------------------------------------------

[m,n] = size(x);
if nargin<4, pconf=[0.95 0.99 0.999]; end
if nargin<3, miss=inf; end
if nargin<2, tf=''; end

% -------------Compute the ICC using Analysis of Variance----------------
ti = zeros(m,1); ni = ti; tj = zeros(1,n); nj = tj;
sx2 = 0;
for i = 1:m
    for j = 1:n
        if x(i,j) ~= miss
            ti(i,1) = ti(i,1)+x(i,j);
            ni(i,1) = ni(i,1)+1;
            tj(1,j) = tj(1,j)+x(i,j);
            nj(1,j) = nj(1,j)+1;
            sx2 = sx2 + x(i,j)^2;
        end
    end
end
mitem = ti./ni;
N = sum(ni); t = sum(ti); ss = sx2 - t^2/N;
ssi = sum(ti.^2./ni) - t^2/N;
ssj = sum((tj.^2./nj),2) - t^2/N;
```

```
ssij = ss - ssi - ssj;
dfi = m-1; dfj = n-1; dfij = N-1-dfi-dfj;
msi = ssi/dfi;
vij = ssij/dfij; vi = max(0,(msi-vij)/n);
qAV = vi/vij; icc = vi/(vi+vij/n); Fobs=msi/vij;         % Main statistics
Q1f=quantF(1-(1-pconf)/2,dfi,dfij);   % Compute the ICC confidence intervals
Q2f=quantF(1-(1-pconf)/2,dfij,dfi);
conf=zeros(length(pconf),3);
for i=1:length(pconf)
    conf(i,1)=pconf(i);
    conf(i,2)=1-Q1f(i)/Fobs;
    conf(i,3)=1-1./(Q2f(i)*Fobs);
end

% ---Validity test using Permutation Resampling for several group sizes----
T=500;                    % Resampling size
[ng0,ngstep,nng]=nppg(n,12); ngs=(1:nng)'*ngstep+ng0;  % Group sizes choice
nrsObs=zeros(nng,3);
for p=1:nng               % Group size loop
ng=ngs(p);                % Number of participants per group
rt = zeros(T,1);
for t = 1:T               % Permutation Resampling loop
    ok = false;
  while ~ok
    xp = x(:,randperm(n)); % Random participant permutation
    ng1 = zeros(m,1); mg1=ng1; ng2=ng1; mg2=ng1;
    for i = 1:m
        for j = 1:ng                        % First group
            if xp(i,j) ~= miss
                ng1(i) = ng1(i)+1; mg1(i) = mg1(i)+xp(i,j);
            end
        end
        for j = (ng+1):(2*ng)               % Second group
            if xp(i,j) ~= miss
                ng2(i) = ng2(i)+1; mg2(i) = mg2(i)+xp(i,j);
            end
        end
    end
    if (min(ng1)>0) && (min(ng2)>0), ok = true; end
  end
  mg1 = mg1./ng1; mg2 = mg2./ng2;
```

```
    rr = corrcoef([mg1, mg2]); rt(t)=rr(1,2);
end
nrsObs(p,:) = [ng,mean(rt),std(rt)];   % [group size, average r, S.D. of r]
end
[q,Chi2]=minChi2(nrsObs,T);         % Estimate optimal q
Chi2df=nng;
Chi2p=1-probChi2(Chi2,Chi2df);      % Validity test type I error risk
rPred=qn2r(q,ngs);                  % Predicted r values for all group sizes
r=qn2r(q,n);                        % Extrapolate r for n participants
if ~strcmp(tf,'')
    tit=plotfit(nrsObs,rPred,Chi2,Chi2df,Chi2p,tf);       % Visualization
end
end

function x = quantF(p,d1,d2)
% F distribution quantiles
x = quantbeta(p,d1/2,d2/2);
x = x.*d2./((1-x).*d1);
end

function x = quantbeta(p,a,b)
% Beta distribution quantiles
tol=1e-6;
x0=zeros(size(p)); x1=ones(size(p));
x=0.5*(x0+x1); dp=betainc(x,a,b)-p;
while max(abs(dp(:)))>tol
    x0(dp<=0)=x(dp<=0); x1(dp>=0)=x(dp>=0);
    x=0.5*(x0+x1); dp=betainc(x,a,b)-p;
end
end

function [q,Chi2]=minChi2(nrs,T)
% Optimize q to minimize Chi^2 (Newton-Raphson method)
tol=1e-9; todo=true; count=0;
s2i=1./nrs(:,3).^2;
q0=s2i'*rn2q(nrs(:,2),nrs(:,1))/sum(s2i);
while todo && (count<30)
    r0=qn2r(q0,nrs(:,1));
    d= T*2*sum((r0/q0).*(nrs(:,2)-r0).*(r0-1)./nrs(:,3).^2);
d2=T*2*sum(((r0/q0).^2).*((r0-1).^2+2*(nrs(:,2)-r0).*(1-r0))./nrs(:,3).^2);
    dq=d/d2; q=q0-dq; count=count+1;
```

```
        if abs(dq)<abs(tol*q)
            todo=false;
        else
            q0=q;
        end
end
Chi2=T*sum(((nrs(:,2)-qn2r(q,nrs(:,1)))./(nrs(:,3))).^2);
if count>=30, warning('Newton-Raphson method failed to converge'); end
end

function q = rn2q(r,n) % Provides q given r and n
q = r./(n.*(1-r));
end

function r = qn2r(q,n) % Provides r given q and n
r = (n.*q)./(n.*q+1);
end

function p = probChi2(x,df)
% Chi-square cumulative probability function
p=gammainc(x/2,df/2);
end

function [S0,S,K]=nppg(n,ek)
% Optimize the number of participants per group in about ek groups
maxng=fix(n/2);
if maxng<=ek
    S0=0; S=1; K=maxng;
else
    minerr=inf;
    for s=1:maxng
        k=fix(maxng/s);
        s0=maxng-s*k;
        err=s0+s*abs(k-ek);
        if err<minerr
            K=k; S=s; S0=s0;
            minerr=err;
        end
    end
end
end
```

```
function tit = plotfit(nrsObs,rPred,Chi2,Chi2df,Chi2p,tf)
% Plot the correlation fit
ns=nrsObs(:,1); r=nrsObs(:,2); s=nrsObs(:,3);
maxx=2*ns(length(ns))-ns(length(ns)-1);
Chi2=round(Chi2*100)/100; Chi2p=round(Chi2p*10000)/10000;
if Chi2p==0, Chi2p=0.0001; end
plot(ns,rPred,'-xk',ns,r(:,1),'--ok');
axis([0 maxx min(0,min(r-s)*1.05) 1]); set(gca,'FontSize',12);
xlabel('Number of participants per group','FontSize',12);
ylabel('r','FontSize',12,'FontWeight','bold');
legend('predicted r','mean observed r (±SD)','Location','Best');
hold on
for i=1:length(ns)
    plot([ns(i);ns(i)],[r(i)-s(i);r(i)+s(i)],'-k');
end
tit=strcat(tf,': \chi^2(',num2str(Chi2df),')=',num2str(Chi2),...
    ', p <',num2str(Chi2p));
title(tit,'FontSize',12);
hold off
end
```

# Appendix B

Example of use of the ECVT program: analysis of the English word naming time database

>> [qAV,icc,conf,r,Chi2,Chi2df,Chi2p] = ECVT(EnglishRT,'English word naming time',0)

*Comment: the input argument EnglishRT is the 770 x 94 data table, the string argument 'English word naming time' is the title for the output figure (Figure 3), the input argument 0 is the code for missing data in the data table. We omit the last input argument (pconf) in order to obtain the 3 default confidence intervals of the ICC (95%, 99%, and 99.9%). We omit the last output argument (mitem) because we do not need the 770 average RT vector. Then, we obtain the output:*

qAV =

   0.1333

*This is the q ratio as it is computed by the ANOVA*

icc =

   0.9261

*This is the ICC as it is computed by the ANOVA*

conf =

   0.9500   0.9185   0.9334
   0.9900   0.9160   0.9355
   0.9990   0.9130   0.9379

*These are the 3 default confidence intervals of the ICC (probability lower_limit upper_limit)*

r =

   0.9236

*This is the ICC as it is estimated by permutation resampling and extrapolation*

Chi2 =

   10.3450

*This is the $\chi^2$ test value (equation 20) for the data set*

Chi2df =

   11

*Number of degrees of freedom of $\chi^2$*

Chi2p =

   0.4996

*Probability of $\chi^2$ under the null hypothesis. Here, the test is not significant, which means that the data in the EnglishRT table fulfil the variable model (1), and the ICC is a valid reference to test item performance models.*

>> print -r600 -dtiff English.tif

*We save the correlation plot figure (Figure 3) which appeared in a separate figure window.*
___________________________________________________________________________

**AUTHOR NOTE**

The authors would like to thank the three anonymous reviewers and the action editor, Dr. Ira Bernstein, for their helpful comments. Part of this study was funded by ERC Research Grant 230313. Correspondence concerning this article should be addressed to P. Courrieu, Laboratoire de Psychologie Cognitive, UMR CNRS 6146, Université de Provence, Centre Saint Charles, Bat. 9, Case D, 3 Place Victor Hugo, 13331 marseille Cedex 3, France. (E-mail: pierre.courrieu@univ-provence.fr)

## Table Legends and Figure Captions

**Table 1.** Observed rejection frequencies of the null difference hypothesis for two $\alpha$ risks (.01 and .05) in the validity test applied to artificial data sets with different $u$ ratios (0, 1/36, 1/16, 1/4). The null hypothesis is true in the case $u=0$ only. In both cases, the data sets had 360 items × 120 participants, with $q=1/16$, and 200 random data sets were tested for each $u$ value.

**Table 2.** Detail of the frequency of abusive detections of under-fits and over-fits at the exact model complexity (20 parameters), using the ICC 99% confidence interval, in the experiments of Figure 6.

**Figure 1.** Predicted and observed mean correlation values (with SD bars) as functions of the number of selected participants per group, in 4 artificial data sets (both with 360 items × 120 participants, $q=1/16$), with different $u$ ratios (0, 1/36, 1/16, 1/4). Predicted and observed functions are not significantly different for $u=0$, but they are significantly different for all non-zero values of $u$, even in those cases where the difference of graphs is just visible.

**Figure 2.** Predicted and observed mean correlation values (with SD bars) as functions of the number of selected participants per group, in the English word identification time data set (120 words × 140 participants, 4 % missing data), from Rey et al. (2009). The two functions are not significantly different ($\chi^2(14)=8.62$, n.s.). The overall ICC is equal to 0.9016, with a 99% confidence interval of [0.8655, 0.9315].

**Figure 3.** Predicted and observed mean correlation values (with SD bars) as functions of the number of selected participants per group, in the English word naming time data set (770 words × 94 participants, 3.61 % missing data). The two functions are not significantly different ($\chi^2(11)=10.35$, n.s.). The overall ICC is equal to 0.9261, with a 99% confidence interval of [0.9160, 0.9355].

**Figure 4.** Predicted and observed mean correlation values (with SD bars) as functions of the number of selected participants per group, in the French word naming time data set (615 words × 100 participants, 3.94 % missing data). The two functions are not significantly

different ($\chi^2(12)=6.60$, n.s.). The overall ICC is equal to 0.9578, with a 99% confidence interval of [0.9513, 0.9638].

**Figure 5.** Variation of $r^2$, and its intersection with the ICC 99% confidence interval, as a function of the number of free parameters used in least-squares fitted models, while original artificial data were generated by a model using exactly 20 parameters (plus random variables), with 61 or 610 items, 40 participants, and two levels of the $q$ ratio. $r^2$ values under the lower ICC confidence limit correspond to under-fitted models, while $r^2$ values above the upper ICC confidence limit correspond to over-fitted models.

**Figure 6.** Detection frequency of under-fits and over-fits, by $r^2$ values outside the ICC 99% confidence interval, as functions of the number of model parameters (complexity) in experiments similar to those of Figure 5, repeated 200 times each. The exact model complexity corresponds to 20 parameters.

**Table 1**

|  | u = 0 | u = 1/36 | u = 1/16 | u = 1/4 |
|---|---|---|---|---|
| α = 0.01 | 0.020 | 0.885 | 1.000 | 1.000 |
| α = 0.05 | 0.040 | 0.960 | 1.000 | 1.000 |

**Table 2**

|  | m=61, q≈1/4 | m=61, q≈1/16 | m=610, q≈1/4 | m=610, q≈1/16 |
|---|---|---|---|---|
| Under-fits | 0.030 | 0.010 | 0.005 | 0.005 |
| Over-fits | 0.005 | 0.005 | 0.005 | 0.005 |
| Total misfits | 0.035 | 0.015 | 0.010 | 0.010 |

**Figure 1**

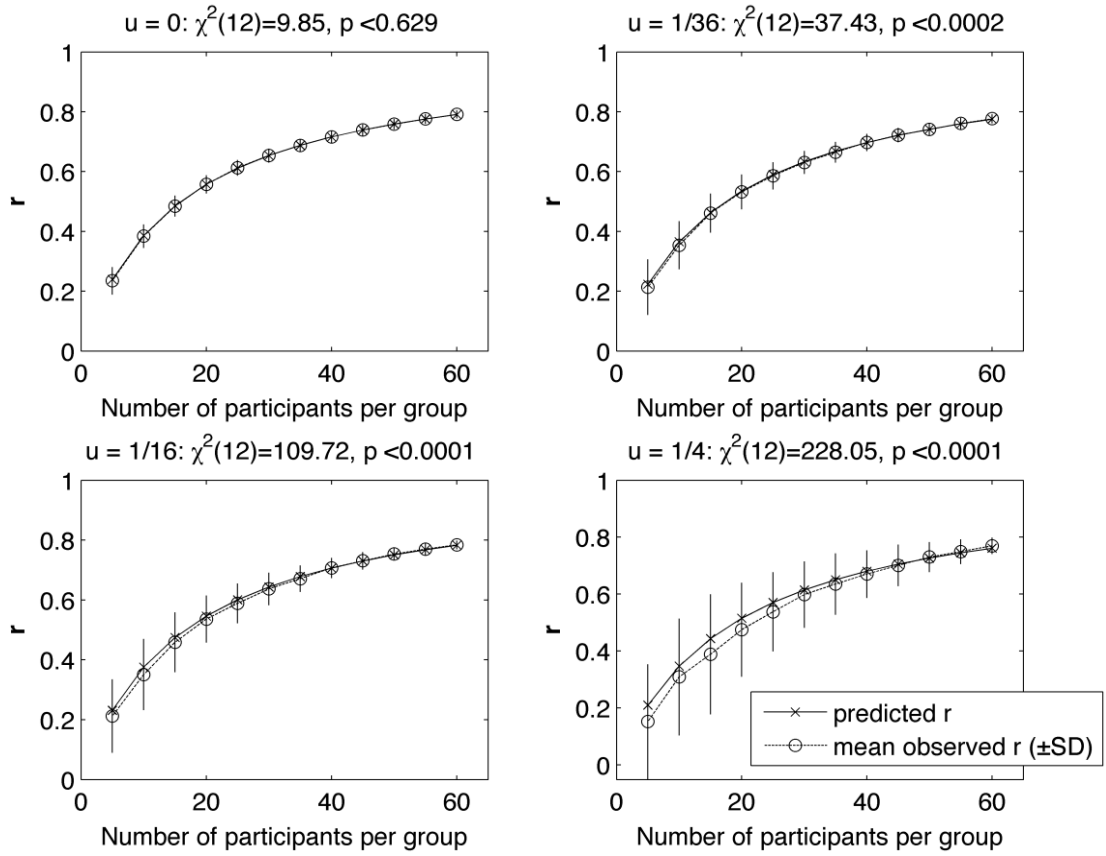

**Figure 2**

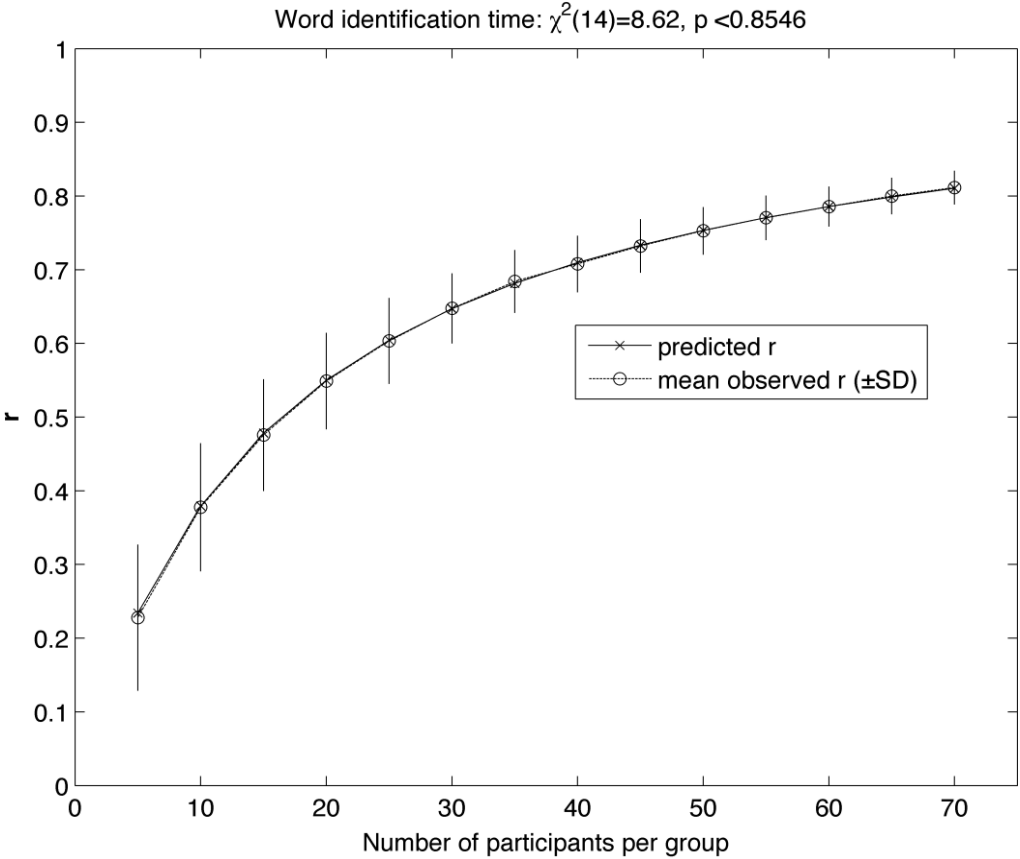

**Figure 3**

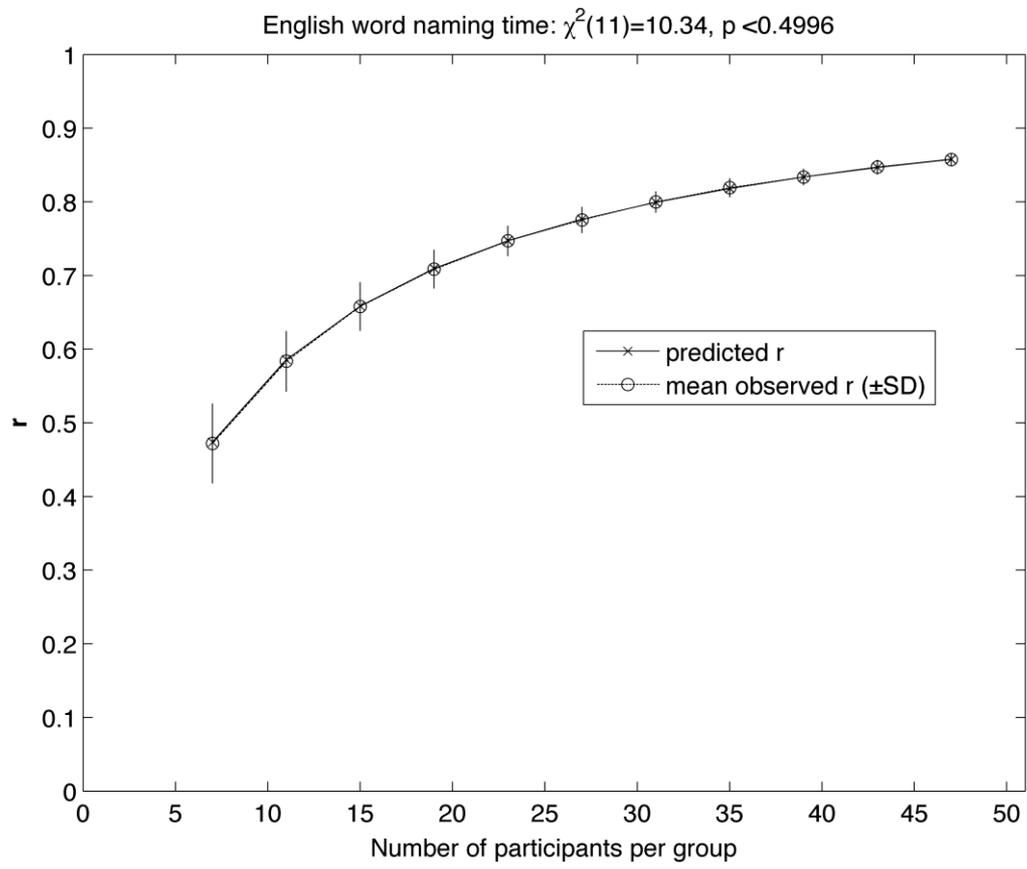

**Figure 4**

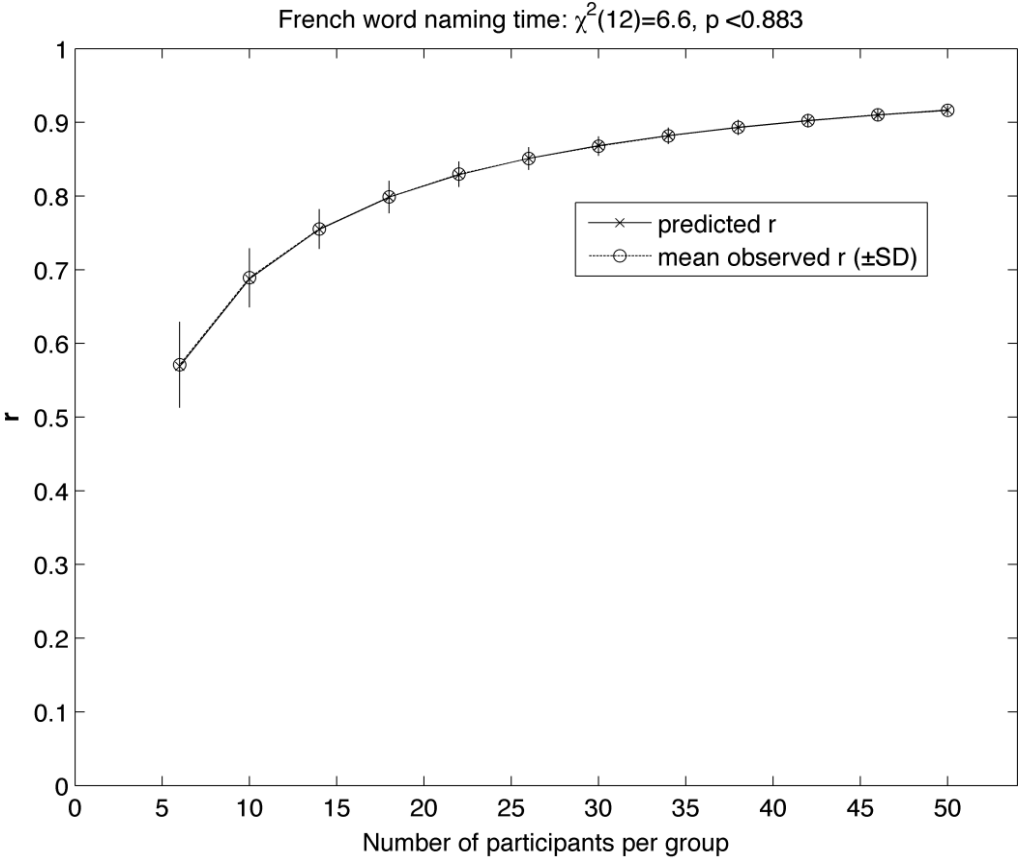

**Figure 5**

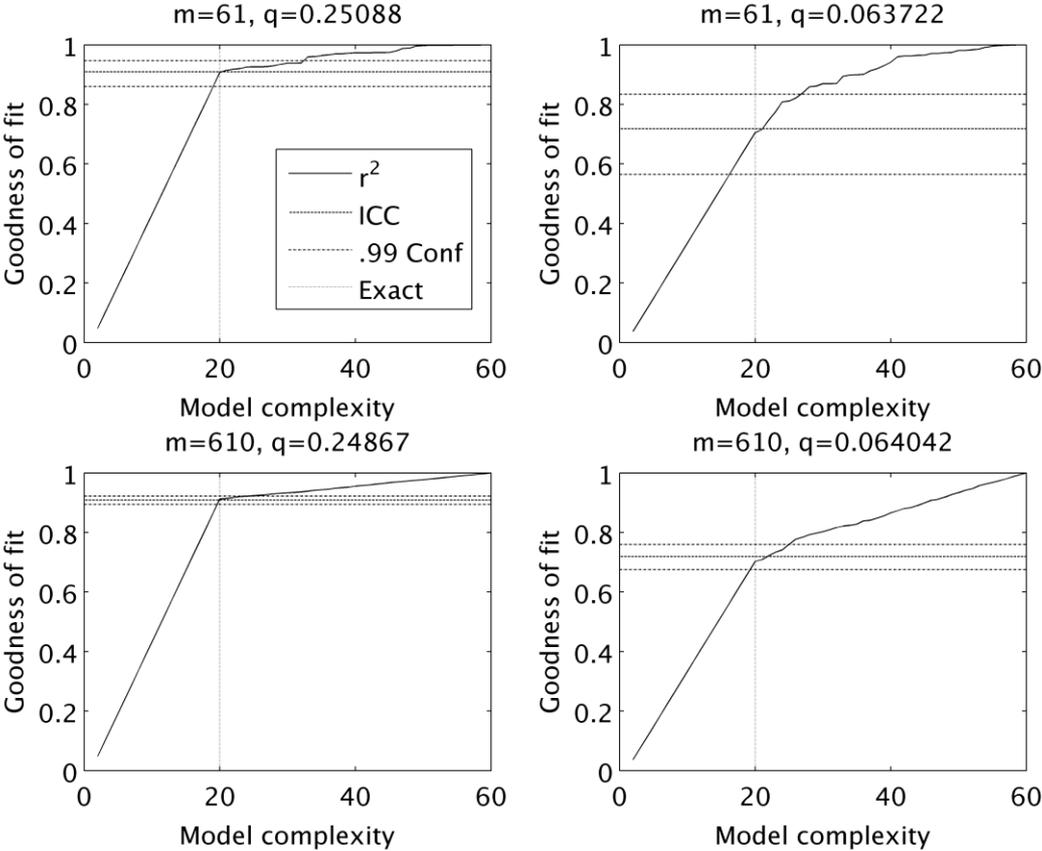

**Figure 6**

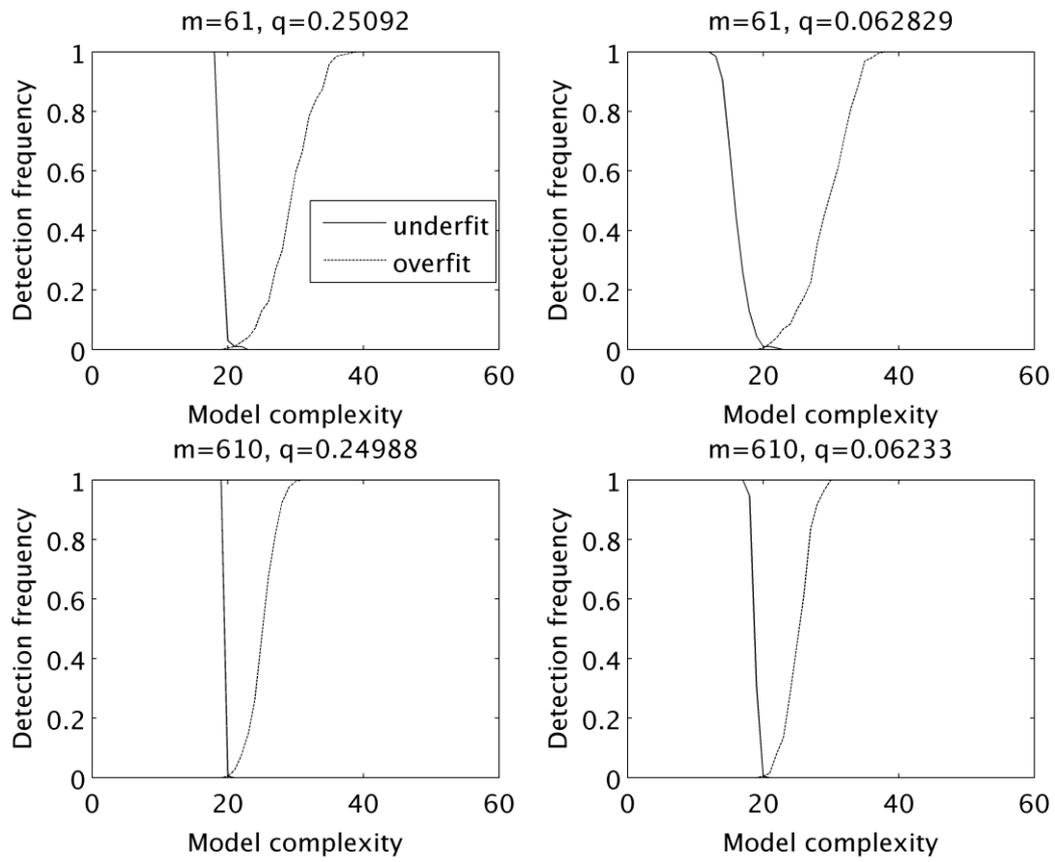